\documentstyle[psfig,epsfig,12pt]{article}

\newcommand{\be}{\begin{equation}}
\newcommand{\ee}{\end{equation}}
\newcommand{\bea}{\begin{eqnarray}}
\newcommand{\eea}{\end{eqnarray}}

\title{On the Existence of the Quantum Action}

\author{
H.~Kr\"{o}ger$\footnote{ 
Email: hkroger@phy.ulaval.ca}$,  
\\ [2mm]
{\small\sl D\'{e}partement de Physique, Universit\'{e} Laval, Qu\'{e}bec, Qu\'{e}bec G1K 7P4, Canada} \\ 
}

\begin{document} 
\maketitle

\noindent Abstract: \\
We have previously proposed a conjecture stating that quantum mechanical transition amplitudes can be parametrized in terms of a quantum action.
Here we give a proof of the conjecture and establish the existance of a local 
quantum action in the case of imaginary time in the Feynman-Kac limit (when temperature goes to zero).
Moreover we discuss some symmetry properties of the quantum action. \\


\setcounter{page}{0}

\newpage

\section{Introduction: Definition and use of the quantum action}
Since the early days of quantum mechanics, many attempts have been made to link quantum mechanics to some concept of classical-like action. First, Wentzel, Kramers, Brillouin proposed the so-called WKB method \cite{WKB}. 
Then there was Bohm's formulation of quantum mechanics \cite{Bohm}.
More modern is the effective action \cite{EffAct} and the Gaussian effective action \cite{GaussEffAct}. Also Gutzwiller's trace formula \cite{Gutz} should be mentioned here, which establishes an approximate expression of the quantum mechanical density of states in terms of classical periodic orbits.  

\bigskip

\noindent The physical reasons, why such a concept is attractive are the following: First of all, quantum mechanics eludes human intuition being shaped by macroscopic physics, i.e. classical physics. Thus a classical-like action in quantum physics is anschaulich. Second, there are concepts playing an important role in modern physics, which have its origin in classical physics. Examples are quantum chaos and quantum instantons. A number of the above approaches have been explored to investigate quantum chaos. Instantons play a role in quantum mechanics: (a)  tunneling and double well potentials (chemical binding, reactions), (b) in high energy physics in the mechanism of quark confinement and the formation of quark-gluon plasma, (c) in cosmology in the inflationary scenario. 
Again classical-like actions have been employed to explore such physics.
Finally, one should mention also the use of an effective action in the theory of supraconductivity.

\bigskip

\noindent Recently, my co-workers and I have proposed a new kind of classical-like action, the quantum action \cite{Jirari:01a,Jirari:01b,Caron:01,Jirari:01c}.
The quantum action has the virtue of having a form as close as possible to the classical action, giving a local expression for quantum transition amplitudes. 
In Refs.\cite{Jirari:01a,Jirari:01b,Caron:01,Jirari:01c} the quantum action has been postulated and also been explored numerically. Numerical studies 
showed in all cases that the quantum action is a good representation of the quantum amplitudes. It allows to give a new unambiguous definition of quantum instantons and quantum chaos, it allows to construct the quantum analogue of classical phase space and to obtain the quantum analogue of Poincar\'e sections and Lyapunov exponents \cite{Jirari:01b,Caron:01}. 

\bigskip

\noindent The purpose of this paper is to give a mathematical proof of existence of the quantum action in imaginary time in the limit of large transition time. This corresponds to thermodynamics in the low-temperature limit (Feynman-Kac limit of the Euclidean path integral).
One should note that the limit of large transition time is not a marginal case but is of central importance in physics: (a) The zero-temperature limit describes the ground state properties of physical systems. (b) Transition time (real time) going to infinity enters in scattering reactions, hence in the S-matrix and cross sections. 
(c) The limit of large times is also involved in non-linear classical dynamics when computing Lyapunov exponents and Poincar\'e sections. Hence this limit plays a role when computing to compute the quantum analogue of Lyapunov exponents and Poincar\'e sections \cite{Caron:01}. 

\bigskip

\noindent What is the quantum action? Let us recall its definition as proposed 
in Ref.\cite{Jirari:01a}. 
We consider the quantum mechanical transition amplitude
\be
G(x_{fi},t_{fi}; x_{in},t_{in}) 
= \langle x_{fi} | e^{-i H (t_{fi}-t_{in})/\hbar} | x_{in} \rangle 
= \left. \int[dx]~ \exp\left[ \frac{i}{\hbar} S[x] \right] \right|_{x_{in},t_{in}}^{x_{fi},t_{fi}} ~ .
\ee

\bigskip

\noindent {\bf Conjecture} \\
For a given classical action 
\be
S[x] = \int dt \frac{m}{2} \dot{x}^{2} - V(x) ~ ,
\ee
there is a quantum action
\be 
\tilde{S}[x] = \int dt \frac{\tilde{m}}{2} \dot{x}^{2} - \tilde{V}(x) ~ ,
\ee
which allows to express the Q.M. transition amplitude by
\bea
\label{DefQuantumAction}
&& G(x_{fi},t_{fi}; x_{in},t_{in}) = \tilde{Z} 
\exp \left[ \frac{i}{\hbar} 
\left. \tilde{\Sigma} \right|_{x_{in},t_{in}}^{x_{fi},t_{fi}} \right] ~ , 
\nonumber \\
&& \left. \tilde{\Sigma} \right|_{x_{in},t_{in}}^{x_{fi},t_{fi}} 
= \left. \tilde{S}[\tilde{x}_{cl}] \right|_{x_{in},t_{in}}^{x_{fi},t_{fi}}  
= \left. \int_{t_{in}}^{t_{fi}} dt ~ \frac{\tilde{m}}{2} \dot{\tilde{x}}_{cl}^{2} - \tilde{V}(\tilde{x}_{cl}) \right|_{x_{in}}^{x_{fi}}  ~ .
\eea
Here $\tilde{x}_ {cl}$ denotes the classical path corresponding to the action $\tilde{S}$ obeying the boundary conditions,
\be
\tilde{x}_{cl}(t=t_{in}) = x_{in}, ~~~ \tilde{x}_{cl}(t=t_{fi}) = x_{fi} ~ .
\ee
This path makes the action $\tilde{S}[\tilde{x}_{cl}]$ minimal. We exclude the occurrence of conjugate points or caustics. 
$\tilde{Z}$ denotes a dimensionful normalisation factor. Eq.(\ref{DefQuantumAction}) is valid with 
the {\em same} action $\tilde{S}$ for all sets of 
boundary positions $x_{fi}$, $x_{in}$ for a given time interval $T=t_{fi}-t_{in}$. 
The parameters of the quantum action depend on the time $T$.  
Any dependence on $x_{fi}, x_{in}$ enters only via the trajectory 
$\tilde{x}_ {cl}$. Likewise, $\tilde{Z}$ depends on the action parameters and $T$, but not on $x_{fi}, x_{in}$.

\bigskip

\noindent If we want to do thermodynamics we need to go over to imaginary time, $t \to -it$. Then the transition amplitude becomes the Euclidean transition amplitude
\be
\label{eq:EuclTransAmpl}
G_{E}(x_{fi},t_{fi}; x_{in},t_{in}) 
= \langle x_{fi} | e^{- H (t_{fi}-t_{in})/\hbar} | x_{in} \rangle 
= \left. \int[dx] ~ \exp \left[-\frac{1}{\hbar} S_{E}[x] \right] \right|_{x_{in},t_{in}}^{x_{fi},t_{fi}} ~ ,
\ee
the classical action becomes the Euclidean action
\be
\label{eq:EuclClassAct}
S_{E}[x] = \int dt \frac{m}{2} \dot{x}^{2} + V(x) ~ ,
\ee
and the quantum action becomes the Euclidean quantum action
\be 
\tilde{S}_{E}[x] = \int dt \frac{\tilde{m}}{2} \dot{x}^{2} + \tilde{V}(x) ~ .
\ee
This allows to express the Euclidean transition amplitude by
\bea
\label{eq:DefEuclQuantAct}
&& G_{E}(x_{fi},t_{fi}; x_{in},t_{in}) = \tilde{Z_{E}} 
\exp \left[ - \frac{1}{\hbar} 
\left. \tilde{\Sigma}_{E} \right|_{x_{in},t_{in}}^{x_{fi},t_{fi}} \right] ~ ,
\nonumber \\
&& \left. \tilde{\Sigma}_{E} \right|_{x_{in},t_{in}}^{x_{fi},t_{fi}} 
= \left. \tilde{S}_{E}[\tilde{x}_{cl}] \right|_{x_{in},t_{in}}^{x_{fi},t_{fi}}  
= \left. \int_{t_{in}}^{t_{fi}} dt ~ \frac{\tilde{m}}{2} \dot{\tilde{x}}_{cl}^{2} + \tilde{V}(\tilde{x}_{cl}) \right|_{x_{in}}^{x_{fi}}  ~ .
\eea

\bigskip

\noindent In order to compute thermodynamic functions from the partition function one has to impose periodic boundary conditions \cite{Kapusta:89}.
$T$ is related to the temperature $\tau$ and the inverse temperature $\beta$ via
\be 
\beta = \frac{1}{k_{B} {\tau}} = T/\hbar ~ .
\ee
In Ref.\cite{Jirari:01b} we have shown that the expectation value of a quantum mechanical observable $O$ at thermodymical equilibrium can be expressed in terms of the Euclidean quantum action along its classical trajectory from $x_{in}$, $\beta_{in}=0$ to $x$, $\beta$.

\section{Proof of existence of the quantum action}
\noindent Consider 1-D throughout (the generalisation to D=2,3 is straightforward). We work in imaginary time in what follows. For simplicity of notation we drop the subscript Euclidean. 
Let us make some assumptions on the potential $V(x)$:
Let $V(x) \geq 0$. Suppose $V(x)$ has a unique minimum.
Let $V(x)$ be a smooth (sufficiently differentiable) function
of $x$ and let $V(x) \to \infty$ when $|x| \to \infty$.
Under those assumptions on the potential, the transition amplitude satisfies the following properties:

\bigskip

\noindent {\bf Proposition 1} \\
For fixed $T$, $G(y,T;x,0)$ has the following properties: \\
(i) It is a real valued, positive function for all $x$, $y$. \\
(ii) It is a symmetric function under exchange $x \leftrightarrow y$. \\
\\
Why is $G(y,T;x,0) \geq 0$? I give two reasons: \\
(a) Physical reason: $G(y,T;x,0)$ is the solution of a diffusion equation describing the motion from $x$ to $y$. This process has a probability interpretation. A probability is positive. 
(b) Mathematical reason: $G(y,T;x,0)$ can be written in terms of a (Wiener) path integral (see Eq.(\ref{eq:EuclTransAmpl})).  
For each path $x(t)$ the weight factor $\exp[ -S[x]/\hbar ] \geq 0$ is 
positive (we have assumed that the classical potential $V(x) \geq 0$ is positive). 
Hence the sum over paths is also positive.
$G(y,T;x,0)$ being real-valued is a  consequence of the fact that 
the weight factor $\exp[ -S[x]/\hbar ]$ of the path integral is real valued.
The second property (ii) follows  from the first property (i) and making the assumption that $H$ is a self adjoint operator.
Next we define a new function $\eta$ to parametrize $G$. In the following we keep $T$ fixed. 

\bigskip

\noindent {\bf Definition} \\
\be
\label{eq:DefEta}
G(y,T;x,0) = G_{0} \exp[-\eta(y,x)] ~ ,
\ee
where $G_{0}$ is some constant (for fixed $T$) which takes care of the 
fact that $G$ has a dimension ($1/L^{D}$). Thus
\be 
\eta(y,x) = - \log[G(y,T;x,0)/G_{0}] ~ .
\ee
Note: The function $\eta$ is well defined, because $G$ is a real, positive function. Via the previous definition, the properties of $G$ translate into the following properties of $\eta$.

\bigskip

\noindent {\bf Proposition 2} \\
(i) $\eta(y,x)$ is a real-valued function for all $x$, $y$. \\
(ii) $\eta(y,x)$ is symmetric under exchange $x \leftrightarrow y$.

\bigskip

\noindent Comparing the parametrisation of $G$ in terms of 
the function $\eta$, Eq(\ref{eq:DefEta}),
with its parametrisation in terms of the quantum action, Eq.(\ref{eq:DefEuclQuantAct}),
this suggests to identify
\bea
\label{eq:Identify}
G_{0} &=& \tilde{Z} ~ , 
\nonumber \\
\eta(y,x) &=& \left. \frac{1}{\hbar} \tilde{S}[\tilde{x}_{cl}] 
\right|_{x,t=0}^{y,t=T} ~ . 
\eea
The idea of the proof is the following.
We assume that the previous identities hold.
Then we analyze its implications. We will end up
in finding an explicit equation for the kinetic term and the potential term of the quantum action.
Then we start at the end and go backwards though the calculation.
This establishes that the quantum action is consistent, and hence proves its existence.

\bigskip

\noindent Identifying $G_{0}=\tilde{Z}$ is possible and trivial because both are constants. Let us identify $\eta$ with $\tilde{\Sigma}$,
\bea
\label{eq:Identification}
\eta(b,a) &=& \frac{1}{\hbar} \left. \tilde{\Sigma} \right|_{a,t=0}^{b,t=T} 
\nonumber \\
&=& \frac{1}{\hbar} \left. \tilde{S}[\tilde{x}_{cl}] \right|_{a,t=0}^{b,t=T} 
\nonumber \\
&=& \frac{1}{\hbar} \left. \int_{0}^{T} dt ~ \frac{\tilde{m}}{2} \dot{\tilde{x}}_{cl}^{2} + \tilde{V}(\tilde{x}_{cl}) \right|_{a,t=0}^{b,t=T} 
~ .
\eea
The question we want to answer is: Can we find a parameter $\tilde{m}$ and a local quantum potential $\tilde{V}$ (e.g. parametrized by polynomial coefficients $\tilde{v}_{k}$, $\tilde{V}(x) = \sum_{k} \tilde{v}_{k} x^{k}$),
such that 
$\eta(b,a) = \frac{1}{\hbar} \left. \tilde{\Sigma} \right|_{a,t=0}^{b,t=T}$
holds for all $a$, $b$ ?

\bigskip

\noindent In order to analyze this question, we proceed by using the property that $\tilde{S}$ is an action and that $\tilde{x}_{cl}$ is the trajectory which makes $\tilde{S}$ extremal.
Let us consider the functional
\be
\label{eq:ActFunctional}
\tilde{S}[x] = \int_{0}^{T} dt ~ \frac{\tilde{m}}{2} \dot{x}^{2} 
+ \tilde{V}(x) ~ ,
\ee
and calculate the variation of the functional to first order (1st order functional derivative). Usually, one keeps initial and final coordinates fixed and varies the path in between. Now we consider the variation of the path, allowing also a variation if initial and final positions. 
Let us denote
\bea
&& \tilde{x}(t) = \tilde{x}_{cl}(t) + \tilde{h}(t) ~ ,
\nonumber \\
&& \tilde{x}_{cl}(t=0) = a, ~  \tilde{x}_{cl}(t=T) = b ~ ,
\nonumber \\
&& \tilde{h}(t=0) = \delta a, ~  \tilde{h}(t=T) = \delta b ~ ,
\nonumber \\
&& \tilde{x}(t=0) = a + \delta a, ~  \tilde{x}(t=T) = b + \delta b ~ .
\eea
Then we compute
\bea
\label{eq:VarAction}
\delta \tilde{S}[\tilde{x}] 
&=& 
\left. \int_{0}^{T} dt ~ 
\frac{\tilde{m}}{2} \left( \dot{\tilde{x}}_{cl} + \dot{\tilde{h}} \right)^{2}  
+ \tilde{V}(\tilde{x}_{cl} + \tilde{h}) \right|_{a + \delta a}^{b + \delta b}
- \left. \int_{0}^{T} dt ~ 
\frac{\tilde{m}}{2} \left( \dot{\tilde{x}}_{cl} \right)^{2}  
+ \tilde{V}(\tilde{x}_{cl}) \right|_{a}^{b}
\nonumber \\
&=& \left. \int_{0}^{T} dt ~ 
\frac{\tilde{m}}{2} \left( 2 \dot{\tilde{x}}_{cl} \dot{\tilde{h}} + \dot{\tilde{h}}^{2} \right) 
+ \frac{d \tilde{V}}{dx}(\tilde{x}_{cl}) \tilde{h} \right|_{\tilde{x}_{cl}=a, \tilde{h}=\delta a}^{\tilde{x}_{cl}=b, \tilde{h}=\delta b} + O(\tilde{h}^{2})
\nonumber \\
&=& \left. \tilde{m} \dot{\tilde{x}}_{cl} \tilde{h} \right|_{0}^{T} 
+ \int_{0}^{T} dt ~ 
(-) \tilde{m} \ddot{\tilde{x}}_{cl} \tilde{h}  
+ \frac{d \tilde{V}}{dx}(\tilde{x}_{cl}) \tilde{h} + O(\tilde{h}^{2})
\nonumber \\
&=& \tilde{m} \dot{\tilde{x}}_{cl}(T) ~ \delta b 
- \tilde{m} \dot{\tilde{x}}_{cl}(0) ~ \delta a 
+ \int_{0}^{T} dt ~ 
\frac{\delta \tilde{S}}{\delta x(t)} \tilde{h}(t) + O(\tilde{h}^{2})
\nonumber \\
&=& \tilde{p}_{cl}(T) ~ \delta b - \tilde{p}_{cl}(0) ~ \delta a  
+ O(\tilde{h}^{2}) ~ ,
\eea
because $\frac{\delta \tilde{S}}{\delta x(t)} = 0$ for $x(t) = \tilde{x}_{cl}(t)$. 
On the other hand, one has
\be
\label{eq:VarEta}
\delta \eta(b,a) = \frac{\partial \eta}{\partial y}(b,a) ~ \delta b + 
\frac{\partial \eta}{\partial x}(b,a) ~ \delta a ~ .
\ee
Comparing Eqs.(\ref{eq:VarAction},\ref{eq:VarEta}) for terms linear in $\delta a$ and $\delta b$, respectively, we find 
\bea
\label{eq:Cond}
\tilde{p}_{cl}(T) &=& \hbar \frac{\partial \eta}{\partial y}(b,a) ~ ,
\nonumber \\
\tilde{p}_{cl}(0) &=& - \hbar \frac{\partial \eta}{\partial x}(b,a) ~ .
\eea
Those are conditions, which are both necessary and sufficient to guarantee that the partial derivatives of the functions 
$\frac{1}{\hbar} ~ \tilde{\Sigma}|_{x}^{y}$ and $\eta(y,x)$ 
coincide for any pair of boundary points $(y,x)$,
\bea 
\label{eq:PartialDeriv}
&& \frac{ \partial }{\partial x} 
\frac{1}{\hbar} \tilde{\Sigma}|_{x}^{y} 
= \frac{ \partial }{\partial x} \eta(y,x) ~ ,  
\nonumber \\
&& \frac{ \partial }{\partial y} 
\frac{1}{\hbar} \tilde{\Sigma}|_{x}^{y} 
= \frac{ \partial }{\partial y} \eta(y,x) ~ .
\eea
Eq.(\ref{eq:PartialDeriv}) implies
\be
\label{eq:FunctIdent}
\frac{1}{\hbar} ~ \tilde{\Sigma}|_{x}^{y} = \eta(y,x) ~ \mbox{modulo a global constant}.
\ee
The global constant can be absorbed into the constants $G_{0}$ and $\tilde{Z}$, respectively,
and this proves Eq.(\ref{eq:Identify}), and hence the existence of the quantum action.

\bigskip

\noindent However, to complete the proof it remains to be shown that the conditions Eq.(\ref{eq:Cond}) can be satisfied. This is not at all obvious from the outset. The terms on the r.h.s. of (\ref{eq:Cond}) stem from the Q.M. transition amplitude
(\ref{eq:EuclTransAmpl}) derived from a classical action (\ref{eq:EuclClassAct}), with mass $m$ and potential $V(x)$.
The terms on the l.h.s. represent the initial and final momenta, corresponding to the trajectory $\tilde{x}_{cl}(t)$. This trajectory is the solution of the Euler-Lagrange equation of motion, which follows from the requirement
$\frac{\delta \tilde{S}}{\delta x(t)} = 0$.
As $\tilde{S}$ depends on the quantum mass parameter $\tilde{m}$ and the 
quantum potential $\tilde{V}(x)$, consequently also the trajectory $\tilde{x}_{cl}(t)$ will depend on $\tilde{m}$ and $\tilde{V}$. The same is true, in particular, for the velocities at the boundaries $\dot{\tilde{x}}_{cl}(0)$ and
$\dot{\tilde{x}}_{cl}(T)$ and hence also for the momenta at the boundaries
$\tilde{p}_{cl}(0)$ and $\tilde{p}_{cl}(T)$.
In other words, requiring that the condition (\ref{eq:Cond}) holds, imposes a constraint on $\tilde{m}$ and $\tilde{V}$. In the following we will show the Eq.(\ref{eq:Cond}) can be satisfied and that this condition guides us to find 
a suitable $\tilde{m}$ and $\tilde{V}$. The guiding principle will be the principle of conserved energy. Once Eq.(\ref{eq:Cond}) having been established, Eq.(\ref{eq:PartialDeriv}) and Eq.(\ref{eq:FunctIdent}) follow.

\section{Construction of quantum action from energy conservation} 
It remains to be shown how to construct a quantum action, such that condition Eq.(\ref{eq:Cond}) is satisfied. We do this by employing the principle of conservation of energy. Any action of the form 
\be
\tilde{S}[x] = \int_{0}^{T} dt ~ \frac{\tilde{m}}{2} \dot{x}^{2} 
+ \tilde{V}(x)  = \int_{0}^{T} dt ~ \tilde{T}_{kin} + \tilde{V} 
\ee
describes a conservative system, i.e., the force is derived from a potential 
and energy is conserved. This means the energy is conserved during the temporal evolution from $t=0$ to $t=T$. In imaginary time, energy conservation reads
\be
- \tilde{T}_{kin} + \tilde{V} = \epsilon = \mbox{const}.
\ee
Now let us choose a (positive) value of the mass parameter $\tilde{m}$. 
Let us look at the energy balance for the trajectory $\tilde{x}_{cl}$ from
$a$ to $b$.
Using Eq.(\ref{eq:Cond}), we find at $t=0$, denoting  
$\tilde{p}_{cl}^{in} \equiv \tilde{p}_{cl}(0)$,
\bea
&& \tilde{T}_{kin} = \frac{(\tilde{p}_{cl}^{in})^{2}}{2 \tilde{m}}, ~
\tilde{V} = \tilde{V}(a) ~ ,
\nonumber \\
&& \epsilon = - \frac{(\tilde{p}_{cl}^{in})^{2}}{2 \tilde{m}} + \tilde{V}(a) ~ .
\eea
Similarly, we find at $t=T$, denoting $\tilde{p}_{cl}^{fi} \equiv \tilde{p}_{cl}(T)$,
\bea
&& \tilde{T}_{kin} = \frac{(\tilde{p}_{cl}^{fi})^{2}}{2 \tilde{m}}, ~ 
\tilde{V} = \tilde{V}(b) ~ ,
\nonumber \\
&& \epsilon = - \frac{(\tilde{p}_{cl}^{fi})^{2}}{2 \tilde{m}} + \tilde{V}(b) ~ .
\eea
Energy conservation implies
\be
- \frac{1}{2 \tilde{m}} (\tilde{p}_{cl}^{in})^{2} + \tilde{V}(a)
=
- \frac{1}{2 \tilde{m}} (\tilde{p}_{cl}^{fi})^{2} + \tilde{V}(b) ~ ,
\ee
or equivalently,
\be
\label{eq:EnerBalance}
\tilde{V}(b) - \tilde{V}(a) =
\frac{1}{2 \tilde{m}} (\tilde{p}_{cl}^{fi})^{2}
- \frac{1}{2 \tilde{m}} (\tilde{p}_{cl}^{in})^{2} ~ .
\ee
We recall from classical mechanics that the r.h.s. represents the work done
(in imaginary time) when the particle moves from $a$ to $b$ 
\be
\tilde{W} =
- \int_{a}^{b} d\tilde{x} ~ \tilde{m} \frac{d^{2}\tilde{x}}{dt^{2}} =
- \int_{0}^{T} dt ~ \tilde{m} \frac{d\tilde{x}}{dt} \frac{d^{2}\tilde{x}}{dt^{2}} =
\frac{1}{2 \tilde{m}} (\tilde{p}_{cl}^{fi})^{2}
- \frac{1}{2 \tilde{m}} (\tilde{p}_{cl}^{in})^{2} ~ .
\ee
Moreover, we recall from classical mechanics that if the work done by a force 
\be
\tilde{W} = \int_{\tilde{C}} d\tilde{x} ~ \tilde{F}(\tilde{x}) 
\ee
is the same for any path $\tilde{C}$ going from $a$ to $b$, or if the work is zero for any closed path, than we know that a potential exists and the system is conservative. 
Thus combining Eq.(\ref{eq:Cond}) and Eq.(\ref{eq:EnerBalance}), we find the following necessary and sufficient condition for the existance of the quantum action: The quantum action exists and is local, if there is a mass $\tilde{m}$ and a local potential $\tilde{V}(x)$, such that 
\be
\label{eq:CondQuantPot}
\frac{2 \tilde{m}}{\hbar^{2}} \left[ \tilde{V}(b) - \tilde{V}(a) \right] 
= \left( \frac{\partial \eta}{\partial y}(b,a) \right)^{2} 
- \left( \frac{\partial \eta}{\partial x}(b,a) \right)^{2} ~~~ \mbox{holds for all} ~ a, b ~ .
\ee
Finally, we should point out that the calculation has yielded a condition for the product
$\tilde{m} \tilde{V}(x)$ 
but not for each of the terms $\tilde{m}$ and $\tilde{V}(x)$ individually.
The reason for this is some underlying symmetry discussed in 
sect. \ref{sec:Invariance} below.

\section{Feynman-Kac limit}
\noindent In the limit $T \to \infty$, or equivalently, 
when temperature goes to zero,
the Feynman-Kac formula holds,
\be 
G(y,T;x,0) \leadsto_{T \to \infty} \langle y | \psi_{gr} \rangle 
e^{-E_{gr}T/\hbar} \langle \psi_{gr} | x \rangle ~ ,
\ee
where $\psi_{gr}$ is the ground state wave function and $E_{gr}$ the ground state energy. Here we make the assumption that the ground state is not degenerate. 
Eq.(\ref{eq:DefEta}) implies
\be
G_{0} e^{-\eta(y,x)} \leadsto_{T \to \infty} \langle y | \psi_{gr} \rangle 
e^{-E_{gr}T/\hbar} \langle \psi_{gr} | x \rangle ~ .
\ee
Taking the logarithm yields
\be
\label{eq:EtaZeroTemp}
- \eta(y,x) + \log G_{0} \leadsto_{T \to \infty} - E_{gr}T/\hbar + \log[\psi_{gr}(y)]
+  \log[\psi_{gr}(x)] ~ .
\ee
From this we compute
\be
\frac{\partial}{\partial y} \eta(y,x)|_{y=b,x=a} \leadsto_{T \to \infty} - \frac{\partial}{\partial y} \left\{\log[\psi_{gr}(y)] + \log[\psi_{gr}(x)] \right\} |_{x=a}^{y=b}
= - \frac{\psi_{gr}'(b)}{\psi_{gr}(b)} ~ .
\ee
Similarly,
\be
\frac{\partial}{\partial x} \eta(y,x)_{y=b,x=a} \leadsto_{T \to \infty} - \frac{\psi_{gr}'(a)}{\psi_{gr}(a)} ~ .
\ee
Then the general condition, Eq.(\ref{eq:CondQuantPot}) becomes
\be 
\label{eq:QuantPot}
\frac{2 \tilde{m}}{\hbar^{2}} [ \tilde{V}(b) - \tilde{V}(a) ] 
\leadsto_{T \to \infty} \left( \frac{\psi_{gr}'(b)}{\psi_{gr}(b)} \right)^{2} 
- \left( \frac{\psi_{gr}'(a)}{\psi_{gr}(a)} \right)^{2} ~ \mbox{for all} ~ 
a, b ~ .
\ee
This means we need to find $\tilde{m}$ and $\tilde{V}(x)$, which
satisfy
\be 
\label{eq:FinalCond}
\frac{2 \tilde{m}}{\hbar^{2}} \left( \tilde{V}(x) - \tilde{V}_{0} \right) 
= \left( \frac{\psi_{gr}'(x)}{\psi_{gr}(x)} \right)^{2} ~ 
\mbox{for all} ~ x ~ . 
\ee
This condition can be satisfied. This establishes the existence of a local quantum action and finishes the proof.

\section{Check of result for harmonic oscillator}
Let us consider the harmonic oscillator in 1-D (in imaginary time).
For the harmonic oscillator, the Q.M. transition amplitude is given by the 
classical action along its classical path. Thus the quantum action should agree with the classical action. This should hold for any temperature $\tau$ or time $T$. In order to check this let us compute   
the quantum potential and hence the quantum action from the condition Eq.(\ref{eq:CondQuantPot}).
The Q.M. transition amplitude reads \cite{Schulman:81}
\be
G(b,T;a,0) = \sqrt{ \frac{ m \omega }{ 2 \pi \hbar \sinh(\omega T) } }
~ \exp \left[ - \frac{ m \omega}{2 \hbar \sinh(\omega T) }
[(b^{2} + a^{2}) \cosh(\omega T) - 2 b a ] \right] ~ ,
\ee
According to Eq.(\ref{eq:DefEta}), we identify
\bea
&& G_{0} = \sqrt{ \frac{ m \omega }{ 2 \pi \hbar \sinh(\omega T) } } ~ ,
\nonumber \\
&& \eta(y,x) = \frac{ m \omega}{2 \hbar \sinh(\omega T) }
[(y^{2} + x^{2}) \cosh(\omega T) - 2 y x ]  ~ .
\eea
Then we compute
\bea
&& \frac{\partial \eta(y,x)}{\partial y} = 
\frac{ m \omega}{ \hbar \sinh(\omega T) }
[ y \cosh(\omega T) - x ] ~ ,
\nonumber \\
&& \frac{\partial \eta(y,x)}{\partial x} = 
\frac{ m \omega}{ \hbar \sinh(\omega T) }
[ x \cosh(\omega T) - y ]  ~ .
\eea
Consequently, we find
\bea
&& \left( \frac{\partial \eta(b,a)}{\partial y} \right)^{2} 
- \left( \frac{\partial \eta(b,a)}{\partial x} \right)^{2} 
=
\left( \frac{ m \omega}{ \hbar \sinh(\omega T) } \right)^{2}
\left[ (b \cosh(\omega T) - a)^{2} - (a \cosh(\omega T) - b)^{2} \right]  
\nonumber \\
&& =  \left( \frac{ m \omega}{ \hbar } \right)^{2}
[ b^{2} - a^{2} ] ~ .
\eea 
Comparing this with Eq.(\ref{eq:CondQuantPot}) yields
\be
\label{eq:HarmOscPot}
\frac{2 \tilde{m}}{\hbar^{2}} \left[ \tilde{V}(b) - \tilde{V}(a) \right] 
= \left( \frac{ m \omega}{ \hbar } \right)^{2}
[ b^{2} - a^{2} ] 
 ~ .
\ee
This is satisfied if we choose
\bea
&& \tilde{m} = m ~ ,
\nonumber \\
&& \tilde{V}(x) = \frac{1}{2} m \omega^{2} x^{2} ~ .
\eea
Thus the quantum potential coincides with the harmonic oscillator potential, i.e. the classical potential and hence the quantum action coincides with the classical action.

\section{Invariance of Q.M. transition amplitude} \label{sec:Invariance}
One may wonder why Eq.(\ref{eq:CondQuantPot}) 
does not specify the quantum potential, but only the combination
$\tilde{m} \tilde{V}(x)$? 
First, one notes that the stationary Schr\"odinger equation for the ground state is invariant (gives the same wave function) under the transformation
\bea
E_{gr} &\to& \alpha E_{gr} ~ ,
\nonumber \\
\hat{V}(x) &\to& \alpha \hat{V}(x) ~ ,
\nonumber \\
m &\to&  m / \alpha ~ .
\eea
Obviously, the following quantity is an invariant under this transformation, 
\be
m \hat{V}(x) \to m \hat{V}(x) ~ .
\ee

\bigskip

\noindent Let us now consider this symmetry in the general case of finite temperature, corresponding to some finite value of time $T$. 
Let us consider the following scale tranformation of the classical mass $m$, the classical potential $V(x)$ and the transition time $T$, where $\alpha$ is some real positive number,
\bea
\label{eq:ClassScaleTrans}
m &\to& m / \alpha ~ ,
\nonumber \\
\hat{V}(x) &\to& \alpha \hat{V}(x) ~ ,
\nonumber \\
T &\to& T / \alpha ~ .
\eea
Then the Q.M. Hamilton operator $\hat{H} = \hat{p}^{2}/2m + \hat{V}(x)$ transforms like
\be
\hat{H} \to \alpha \hat{H} ~ .
\ee
Because the Q.M. transition amplitude $G$ is a matrix element of an operator-valued function of $\hat{H} T/ \hbar$, being an invariant under the above scale transformation, consequently the Q.M. transition amplitude 
is an invariant also,
\be
\label{eq:InvTransAmpl}
G(b,T;a,0) \to G(b,T;a,0) ~ .
\ee

\bigskip

\noindent Let us now look at invariance properties of the classical system.
Consider the Lagrangian
\be
L(x(t),\dot{x}(t)) = \frac{m}{2} \dot{x}^{2} + V(x) ~ ,
\ee
and the action
\be
S[x] = \int_{0}^{T} dt ~ L(x(t),\dot{x}(t)) ~ .
\ee
The Euler-Lagrange equation of motion reads
\be
- m \ddot{x}_{cl}(t) + \frac{d V(x)}{dx}|_{x = x_{cl}(t)} = 0 ~ ,
\ee
where $x_{cl}(t)$ denotes the solution corresponding to a given pair of boundary points $x_{cl}(t=0)=a$, $x_{cl}(t=T)=b$.
A straight forward computation yields the following transformation rules: \\
(i) Classical trajectory: \\
\be
x_{cl}(t) \to x'_{cl}(t) = x_{cl}(\alpha t) ~ .
\ee
(ii) 
Lagrangian evaluated at classical trajectory: \\
\be
L(x_{cl}(t),\dot{x}_{cl}(t)) \to L'(x'_{cl}(t),\dot{x}'_{cl}(t))
= \alpha L(x_{cl}(\alpha t),\dot{x}_{cl}(\alpha t)) ~ .
\ee
(iii) Action evaluated along classical trajectory: \\
\be
\label{eq:InvClassAction}
S[x_{cl}] \to S'[x'_{cl}] = S[x_{cl}] ~ .
\ee
Thus we see that $\Sigma = S[x_{cl}]$ is an invariant in classical mechanics. Trivially, also $m V(x)$ is an invariant.

\bigskip

\noindent The invariance properties of the classical system immediately carry over to the quantum action. Consider the transformation
\bea
\label{eq:QuantScaleTrans}
\tilde{m} &\to& \tilde{m} / \alpha ~ ,
\nonumber \\
\tilde{V}(x) &\to& \alpha \tilde{V}(x) ~ ,
\nonumber \\
T &\to&  T / \alpha ~ .
\eea
Consequently, $\tilde{\Sigma}$ is an invariant and $\tilde{m} \tilde{V}(x)$ is also an invariant. Under the combined scale transformations, Eqs.(\ref{eq:ClassScaleTrans},\ref{eq:QuantScaleTrans}), we have shown that both, the Q.M. transition amplitude $G$ and the quantum action $\tilde{\Sigma}$ are invariants.    
The lesson from this is that for a given fixed time (corresponding to finite temperature), Eq.({\ref{eq:CondQuantPot}) is not sufficient to determine the quantum potential, but one needs an independent determination of $\tilde{m}$. One way to do this is via use of the renormalisation group equation proposed in Ref.\cite{Jirari:01c}. 
In retrospective, one may consider the invariance properties
Eqs.(\ref{eq:InvTransAmpl},\ref{eq:InvClassAction}) as a hint on a relation between the Q.M. transition amplitude and the quantum action.

\section{Concluding remarks}
The proof is non-perturbative. It does not require the system to be integrable.
The proof can without difficulty be generalized to 3-D (or higher dimensions).
An interesting observation is the following: After the back transformation to real time, the corner stone equation of the proof, Eq.(\ref{eq:Cond}) reads 
\bea
\tilde{p}_{cl}(T) &=& i \hbar \frac{\partial \eta}{\partial y}(b,a) ~ ,
\nonumber \\
\tilde{p}_{cl}(0) &=& - i \hbar \frac{\partial \eta}{\partial x}(b,a) ~ ,
\eea
which relates the classical momentum $\tilde{p}_{cl}$ to the Q.M. momentum operator, $\hat{P}_{x} = i \hbar \partial /\partial x$, which reminds us of canonical quantization rules. 

\bigskip

\noindent {\bf Acknowledgements} \\ 
H.K. is grateful for support by NSERC Canada. 
For discussions, critical reading and constructive suggestions H.K. is very grateful to W. Hengartner, T. Ransford, and L.S. Schulman.

\end{document}